\documentclass[aps,prl,superscriptaddress,nofootinbib]{revtex4-2} 

\usepackage{chemformula} 
\usepackage{longtable} 
\usepackage{multirow} 
\usepackage{amssymb} 
\usepackage{enumitem} 
\usepackage{float} 
\usepackage{pdfpages} 
\usepackage{pgffor} 
\usepackage{tabularx}
\makeatletter
\AtBeginDocument{\let\LS@rot\@undefined}
\makeatother

\begin{document}

\title{Hydrogen Transport Between Layers of Transition Metal-Di\-chal\-co\-ge\-nides}

\author{Ismail Eren}
\affiliation{Helmholtz-Zentrum Dresden-Rossendorf, Abteilung Ressourcenökologie, Forschungsstelle Leipzig, Permoserstr. 15, 04318 Leipzig, Germany}
\author{Yun An}
\affiliation{Beijing Key Laboratory of Theory and Technology for Advanced Batteries Materials, School of Materials Science and Engineering, Peking University, Beijing 100871, China}
\author{Agnieszka B. Kuc}
\email{a.kuc@hzdr.de}
\affiliation{Helmholtz-Zentrum Dresden-Rossendorf, Abteilung Ressourcenökologie, Forschungsstelle Leipzig, Permoserstr. 15, 04318 Leipzig, Germany}

\begin{abstract}

Hydrogen is a crucial source of green energy and has been extensively studied for its potential usage in fuel cells.
The advent of two-dimensional crystals (2DCs) has taken hydrogen research to new heights, enabling it to tunnel through layers of 2DCs or be transported within voids between the layers, as demonstrated in recent experiments by Geim's group.
In this study, we investigate how the composition and stacking of transition-metal dichalcogenide (TMDC) layers influence the transport and self-diffusion coefficients ($D$) of hydrogen atoms using well-tempered metadynamics simulations.
Our findings show that modifying either the transition metal or the chalcogen atoms significantly affects the free energy barriers ($\Delta F$) and, consequently, the self-diffusion of hydrogen atoms between the 2DC layers.
In the $H^h_h$ polytype (2$H$ stacking), MoSe$_2$ exhibits the lowest $\Delta F$, while WS$_2$ has the highest, resulting in the largest $D$ for the former system. Additionally, hydrogen atoms inside the $R_h^M$ (or 3$R$) polytype encounter more than twice lower energy barriers and, thus, much higher diffusivity compared to those within the most stable $H^h_h$ stacking.
These findings are particularly significant when investigating twisted layers or homo- or heterostructures, as different stacking areas may dominate over others, potentially leading to directional transport and interesting materials for ion or atom sieving.

\end{abstract}

\maketitle

\section{\label{sec:intro}Introduction}

Van der Waals (vdW) materials, such as transition-metal dichalcogenides (TMDCs), hexagonal boron nitride ($h$BN), and graphite, to name just a few, provide interstitial voids - empty spaces between their layers.
These voids are typically at least 3~\AA\ in height and enable the transportation of small atoms or ions under such confined conditions.
The growing interest and focus on vdW materials in recent years stem from their potential applications as transport channels for ion-, atom- or isotope-sieving technologies and energy-related applications.\cite{Wang2022review, Wang2023review, ma2023review}

The intercalation of alkali ions (Li$^+$, Na$^+$, K$^+$) between layers of TMDCs has been the subject of both experimental and theoretical investigations for several years, particularly in the context of energy storage and batteries.\cite{choi2017recent, Xu2017, chen2020, Huy2021}
Ion-intercalation is reversible, however, due to the size of alkali ions, it leads to interlayer expansion, affecting not only the electronic properties but also the local polytype, such as a change from 2$H$ to 1$T$.\cite{kappera2014, Shuai2016, Xu2017}
Hydrogen, in both atomic and ionic forms, is one of the smallest species capable of intercalating between TMDC layers, while maintaining the structure of the host material intact.
As a green energy source, hydrogen has been extensively studied, particularly for its potential applications in fuel cells, isotope separation, or catalytic processes, where its transport plays a crucial role.\cite{hu2018transport, an2019chemistry, Kuznetsov2022}

One of the most notable experiments involving hydrogen and layered materials in the last five years was reported by Geim's group in 2018.\cite{hu2018transport}
The authors demonstrated that vdW gaps between 2DC layers provide ångström-size channels, enabling quantum confinement of protons even at room temperature.
Moreover, they observed that protons face higher energy barriers than deuterons while entering these gaps in $h$BN and 2$H$ MoS$_2$.
However, the report did not clarify whether ionic or atomic hydrogen isotopes were being transported, and the precise mechanism of this transport remained unknown.

In 2019, we conducted an investigation to address these remaining questions regarding the transport of protons and protiums between layers of $h$BN and 2$H$ MoS$_2$ materials using quantum mechanics.\cite{an2019chemistry}
To achieve this, we employed well-tempered metadynamics simulations (WTMetaD).
Our findings unveiled that: i) protiums are transported easier than protons and ii) both species exhibit a zigzag path during transport, wherein they hop between two adjacent layers.
We observed that the likelihood of this hopping behavior was linked to the presence of interlayer shear modes, characteristic to all layered materials.

In the present work, we extended our previous studies and investigated the self-diffusion of H atoms between layers of Group-6 TMDC bulk materials with different layer stackings.
Our focus was on examining how the layer composition (metal atoms - Mo, W, Nb, or chalcogen atoms - S, Se) and the layer stacking in MoS$_2$ ($H$- or $R$-types of high-symmetry stackings) influence the free-energy barriers ($\Delta F$) for H transport and the corresponding diffusivity.
To investigate the movement of H atoms between layers and construct the free-energy (F) landscape, we employed WTMetaD simulations.
Our findings revealed that Se-based materials, as well as Mo-based materials, exhibit lower $\Delta F$ and higher self-diffusion coefficients ($D$) compared to their S- or W-based counterparts, making MoSe$_2$ systems ideal for H atom transport.
Among different high-symmetry stackings, the $R$ types have up to twice lower $\Delta F$ than the $H$ types.
These results are of significance in the exploration of 2D TMDC materials as potential transport channels for energy devices, such as fuel cells.
In particular, twisted homo- and heterostructures will be of great interest, as differently sized domains of the mentioned high-symmetry stackings, with varying $\Delta F$, will lead to local transport channels.

\section{\label{sec:results}Results and discussion}
\subsection{\label{sec:stoic}Effect of Elemental Composition}

Initially, we examined the diffusion of H-atoms between layers of various Group-6 TMDC bulk materials in their $H_{h}^{h}$ (2$H$) forms as depicted in Figure~\ref{fig1}.
In addition, we also investigated metallic NbS$_2$ (in $H_{h}^{X}$ as the most stable polytype) and semiconductor main-group $\beta$-InSe for comparison.
Our primary objective was to comprehend the influence of different elements in the layers on $\Delta F$ for hydrogen transfer and its diffusivity.

\begin{figure*}[ht!]
 \includegraphics[width=1\textwidth]{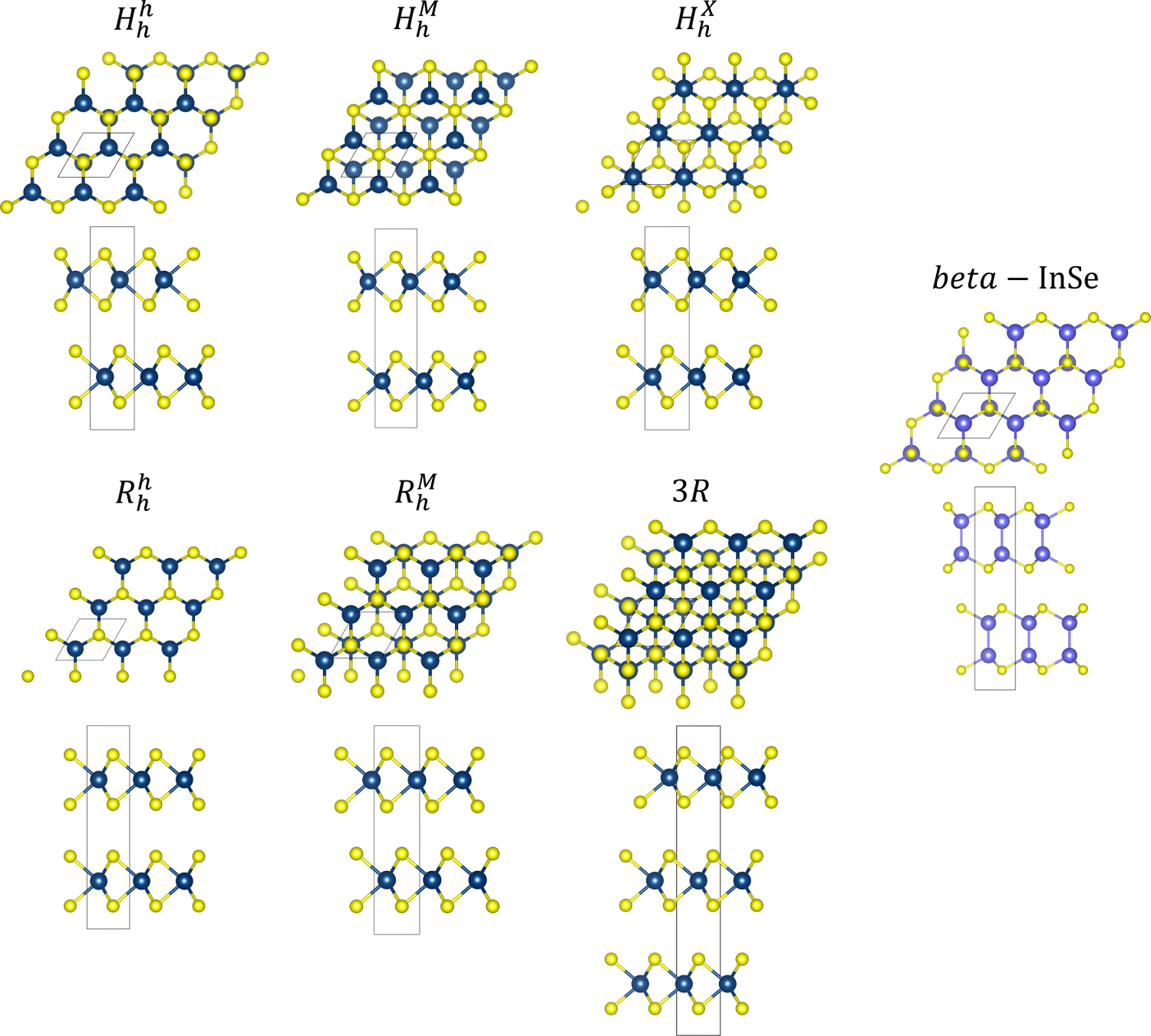}%
 \caption{\label{fig1} 
 Top and side views of all considered structural high-symmetry stackings. Unit cells are marked with black lines. Blue - metal atom (Mo, W, Nb), yellow - chalcogen atom (S or Se), violet - In atom. 
 }
\end{figure*}

The optimized lattice parameters and the interatomic distances of all systems investigated in this section are given in Table~\ref{tab:table1}.
The in-plane lattice parameters agree within a maximum of 0.6\% with the experimental data, while somewhat larger errors were obtained for the out-of-plane lattice parameters, with maximum overestimation of 2.4\% for MoSe$_2$.\cite{Hulliger1976}
All systems have approximately 3~\AA\ of space between the layers ($d_{I}$; see Table~\ref{tab:table1}), providing sufficient room to accommodate H atoms.
The H atoms bind to the chalcogen atoms ($X$) with bond lengths of around 1.4~\AA\ for S and approximately 1.5-1.6~\AA\ for Se.
In all cases, the H atoms bound to $X$ atoms point towards the interlayer void and the middle of hexagonal rings.

\begin{table}[ht!]
 \caption{\label{tab:table1} Structural properties of all investigated bulk materials in the $H_{h}^{h}$ forms ($H_{h}^{X}$ for NbS$_2$): lattice vectors,\textit{a}, \textit{b}, and \textit{c} (in \AA); bond lengths between metal and chalcogen atoms, $d_{M-X}$, and between chalcogen atoms within a single layer, $d_{X-X}$, (in \AA); interlayer distances measured between metal atoms, $d_{M-M}$, and between inner chalcogen atoms in the neighbouring layers, $d_{I}$, (in \AA); and the bond lengths between chalcogen and hydrogen atoms, $d_{X-H}$, (in \AA), cf. Figure~\ref{fig2}b. For InSe, $d_{M-M}$ was measured between inner In atoms.
 Free energy barriers, $\Delta F$, (in meV) and the self-diffusion coefficients, $D$, (in cm$^{2}$ s$^{-1} \times 10^{-3}$) of all studied systems.}
 \begin{ruledtabular}
 \begin{tabular}{lcccccc}
 \hline
  Property & $H_{h}^{h}$-MoS$_{2}$ & $H_{h}^{h}$-MoSe$_{2}$ & $H_{h}^{h}$-WS$_{2}$ & $H_{h}^{h}$-WSe$_{2}$ & $H_{h}^{X}$-NbS$_{2}$ & $H_{h}^{h}$($\beta$)-InSe\\
  \hline
  \textit{a} = \textit{b} & 3.164 & 3.302 & 3.164 & 3.296 & 3.329 & 4.068\\
  \textit{c} & 12.314 & 13.224 & 12.440 & 13.236 & 12.130 & 17.010\\
  $d_{X-M}$  & 2.405 & 2.538 & 2.409  & 2.540 & 2.483 & 2.674 \\
  $d_{X-X}$ & 3.131 & 3.350 & 3.139 & 3.369 & 3.140 & 5.359 \\
  $d_{M-M}$ & 6.159 & 6.620 & 6.226 & 6.617 & 6.050 & 5.703 \\
  $d_{I}$ & 3.030 & 3.262 & 3.083 & 3.251 & 2.918 & 3.143 \\
  $d_{X-H}$  & 1.433 & 1.574 & 1.432 & 1.663 & 1.367 & 1.493 \\
  $\Delta F$  & 115  & 45 & 127 & 68 & 145 & 320\\
  $D$   &  0.27 & 4.15 & 0.18 & 1.70 & 0.09 & 0.11 $\times 10^{-3}$\\
  \hline
 \end{tabular}
 \end{ruledtabular}
\end{table}

\begin{figure}[ht!]
 \includegraphics[width=0.75\columnwidth]{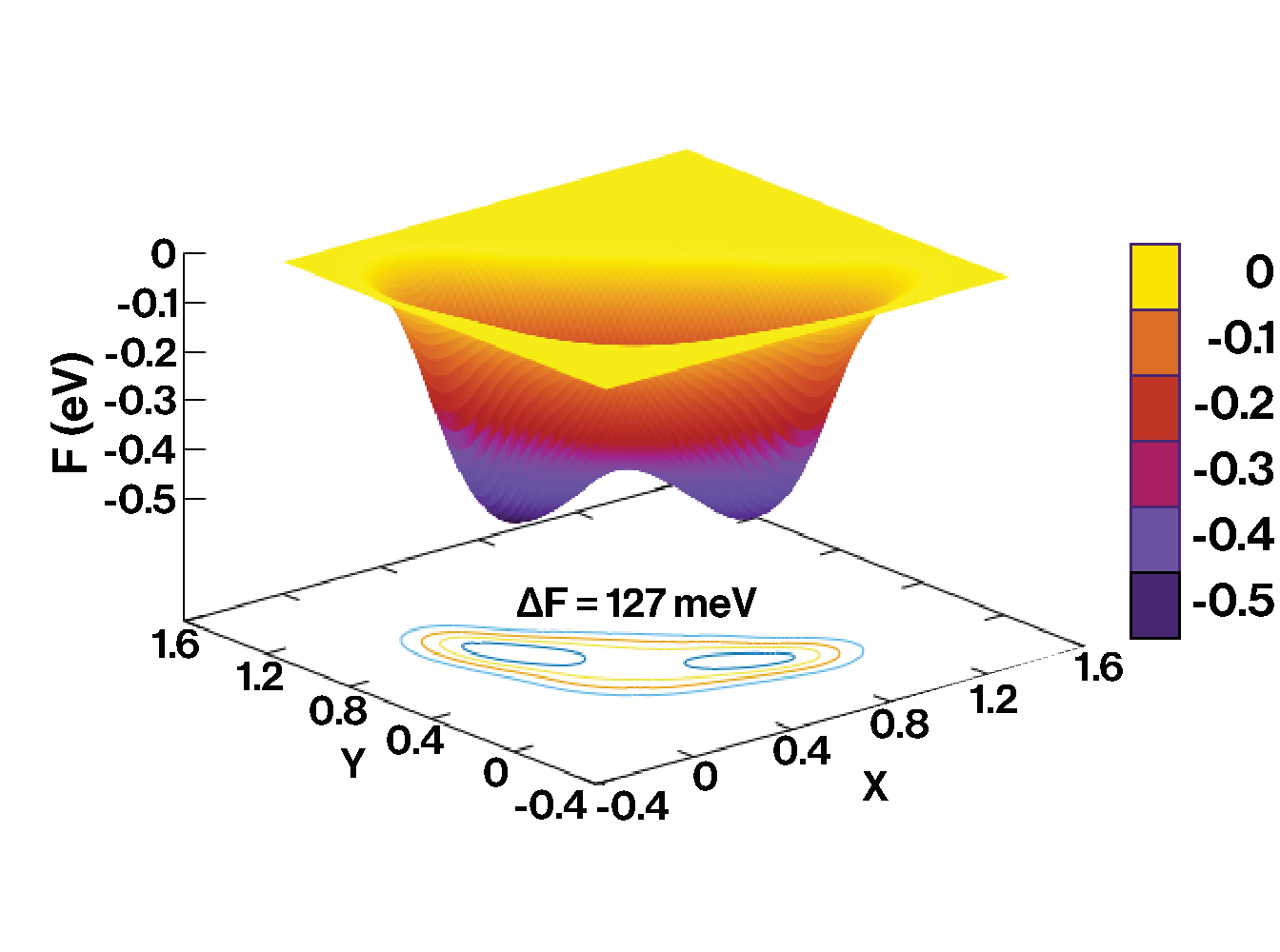}%
 \caption{\label{fig3} 
Exemplary free-energy ($F$) landscape obtained from WTMetaD simulations of a single H atom moving between layers of $H_{h}^{h}$ bulk WS$_{2}$. The collective variables (CVs) were defined as coordination number (CN) of H atoms with the inner chalcogen atoms X1-H (CV1) and X2-H (CV2; cf. Figure~\ref{fig2}), which correspond to the x and y axes. The color legend corresponds to $F$ in eV.}
\end{figure}

We employed WTMetaD simulations, with two collective variables (CVs) defined for each system (see Section~\ref{sec:methodology} for detailed information), to calculate the free-energy ($F$) landscape.
The landscape exhibited two minima and an energy barrier ($\Delta F$) separating them.
As an example, Figure~\ref{fig3} displays the $F$ landscape for bulk WS$_2$ along with its corresponding $\Delta F$ value.
The energy landscapes for all the other systems share similar characteristics and primarily vary in terms of their $\Delta F$ values.
For completeness, these energy landscapes are provided in Figure~S1 in the Supporting Information.
The values of $\Delta F$ for all systems are given in Table~\ref{tab:table1}.

Among the four Group-6 TMDC bulk materials, the highest $\Delta F$ belongs to WS$_2$ and the lowest to MoSe$_2$.
We noticed that Se-based TMDC have lower $\Delta F$ than their S-based counterparts and Mo-based systems have lower $\Delta F$ than the corresponding W-based materials.
From Equation~\ref{eq2} in Section~\ref{sec:methodology} we know that lower the energy barrier the larger the self-diffusion coefficient.
This allows a conclusion that H atoms diffuse easier in the Se- and Mo-based TMDCs.
The significance of interlayer spacing cannot be understated, as it plays a pivotal role in self-diffusion.
Our simulations, however, tend to overestimate the interlayer distance in Se-based systems to a greater extent compared to S-based materials.
To address this, we extended our investigation on MoSe$_2$ bulk and adjusted the $c$ lattice vector to match the experimental value (12.910~\AA).\cite{Hulliger1976}
By doing so, the H-binding sites are brought into closer proximity, reducing the $\Delta F$ even further.
Consequently, the self-diffusion of MoSe$_2$ increases, supporting our conclusions.

Our results agree well with previously published work,\cite{an2019chemistry} where the authors obtained $\Delta F$ = 90~meV and $D$ = 1.50 $\times$ 10$^{-4}$ cm$^2$ s$^{-1}$ for MoS$_2$ bulk in $H_{h}^{h}$ form.
The small differences that are observed might be due to the 0.4\% difference in the interlayer distance.

It is important to note that the defined collective variables (CVs) in WTMetaD simulations do not allow for the investigation of diffusion direction; they only result in a Brownian motion of H atoms. Nevertheless, these CVs enable the estimation of $D$ (self-diffusion coefficients).

In agreement with our previous report,\cite{an2019chemistry} the H atoms are transported between the layers of TMDC materials in a zigzag manner, binding to the inner chalcogen atoms of neighbouring layers (see Figure~S2).
The transfer between layers is also supported by low-energy shear modes.
Additionally, our previous results shows that energy barrier for proton transport between the layers is only slightly higher than that for H atom, which should be similar for other TMDCs.
This might be important when investigating proton transport for potential applications, such as fuel cells.

For comparison, we investigated two other layered materials, NbS$_2$ and $\beta$-InSe.
NbS$_2$ is a metallic TMDC from Group-5.
Nb atoms have one electron less than Mo atoms, what results in a ferromagnetic material.
This system was investigated in the $H_{h}^{X}$ polytype, which is more stable than the $H_{h}^{h}$ one.\cite{Jellinek1960MolybdenumAN}
The interlayer distance in this material is smaller than in the Group-6 TMDCs, however, $\Delta F$ is by almost 20~meV higher than that of WS$_2$, resulting in a very low $D$ (only 9 $\times$ 10$^{-5}$ cm$^2$ s$^{-1}$).
The second material, $\beta$-InSe ($H_{h}^{h}$ stacking), is a well-studied layered system, which is a quasi-direct band-gap semiconductor.\cite{eren2019vertical}
It has the highest $\Delta F$ of all studied systems, over 300~meV, which is more than twice larger than that of NbS$_2$, resulting in a negligible self diffusion (1.1 $\times$ 10$^{-7}$ cm$^2$ s$^{-1}$).
This might be due to the fact that H atom forms much shorter bonds with Se in this material than in the Group-6 TMDCs, which should result in different frequency of vibration ($v_{0}$).

\subsection{\label{sec:stacks}Effect of Stacking}

In this section, we investigated only bulk MoS$_{2}$ in different high-symmetry structural stacking, namely $H_{h}^{h}$, $H_{h}^{M}$, $H_{h}^{X}$, $R_{h}^{h}$, and $R_{h}^{M}$ (which is the same as $R_{h}^{X}$ in homo-layered materials; see Figure~\ref{fig1}).\cite{yu2018brightened,yuan2020twist}
Additionally, we considered the 3$R$ stacking, which requires three layers in the bulk unit cell.

Among the five high-symmetry stackings, there are three that are low-energy systems:\cite{arnold2023} the most stable $H_{h}^{h}$, and the $R_{h}^{M}$ and $H_{h}^{X}$, which are only about 2~meV per unit cell less stable.
The other two stackings are about 53~meV per unit cell less stable than $H_{h}^{h}$.
It, therefore, was not surprising that, during the WTMetaD simulations with H atoms between the layers, the high-energy stackings ($R_{h}^{h}$ and $H_{h}^{M}$) reverted back to their corresponding $R_{h}^{M}$ and $H_{h}^{X}$ forms.
Consequently, Table~\ref{tab:table2} presents the results for the most stable structural stackings of MoS$_2$.
The corresponding free-energy landscapes are shown in Figure~S1.

\begin{table}[ht!]
 \caption{\label{tab:table2} Structural properties of MoS$_2$ bulk systems with different high-symmetry stackings: lattice vectors,\textit{a}, \textit{b}, and \textit{c} (in \AA); bond lengths between metal and chalcogen atoms, $d_{M-X}$, and between chalcogen atoms within a single layer, $d_{X-X}$, (in \AA); interlayer distances measured between metal atoms, $d_{M-M}$, and between inner chalcogen atoms in the neighbouring layers, $d_{I}$, (in \AA); and the bond lengths between chalcogen and hydrogen atoms, $d_{X-H}$, (in \AA), cf. Figure~\ref{fig2}b.
 Free energy barriers, $\Delta F$, (in meV) and the self-diffusion coefficients, $D$, (in cm$^{2}$ s$^{-1} \times 10^{-3}$) of all studied systems..
 }
 \begin{ruledtabular}
 \begin{tabular}{lcccc}
 \hline
  Property &
  $H_{h}^{h}$-MoS$_{2}$ & $H_{h}^{X}$-MoS$_{2}$ & $R_{h}^{M}$-MoS$_{2}$ & 3$R$-MoS$_{2}$\\
  \hline
  \textit{a} = \textit{b} & 3.164 & 3.167 & 3.166 & 3.167 \\
  \textit{c} & 12.314 & 12.311 & 12.191 & 18.194 \\
  $d_{X-M}$  & 2.405 & 2.406 & 2.406 & 2.405 \\
  $d_{X-X}$  & 3.131 & 3.127 & 3.130 & 3.125 \\
  $d_{M-M}$  & 6.159 & 6.153 & 6.095 & 6.069 \\
  $d_{I}$    & 3.030 & 3.025 & 2.967 & 2.942 \\
  $d_{X-H}$  & 1.433 & 1.459 & 1.462 & 1.460 \\
  $\Delta F$  & 115  & 80 & 50 & 61 \\
  $D$  &  0.27 & 1.06 & 3.30 & 2.13 \\
  \hline
 \end{tabular}
 \end{ruledtabular}
\end{table}

While the phononic properties of the $H_{h}^{h}$ and 3$R$ MoS$_2$ polytypes show very similar frequency for shear modes (31.85 $cm^{-1}$ and 32.85 $cm^{-1}$ for 3$R$- and $H_{h}^{h}$ polytypes, respectively),\cite{van2019stacking} these materials differ significantly in the diffusivity of H atoms.
The $H_{h}^{h}$ polytype has the highest $\Delta F$ and the lowest $D$ among all the stackings.
On the other hand, the lowest energy barrier and the highest diffusion belong to the $R$ forms.
We expect similar results for the other investigated TMDCs.

This finding holds significant importance when considering twisted bilayer TMDCs with moiré structures and high-symmetry stacking domains\cite{arnold2023} for ion- or atom-transporting materials in energy technologies.
Depending on the twist angle between layers, different sizes of the mentioned high-symmetry stackings are formed, which may lead to localized transport channels and directional diffusion (a subject of our ongoing investigations).

\section{\label{sec:conclusions}Conclusions}

In summary, our study focused on investigating the diffusivity of H atoms between layers of TMDC bulk materials, utilizing the interstitial voids that provide ample space for H-atom or -ion transport.
We observed that Group-6 TMDCs containing Se or Mo atoms exhibit lower free-energy barriers for H-atom transport compared to their S- or W-based counterparts.
Additionally, we explored the impact of layer stacking in MoS$_2$ on the self-diffusion coefficients.
Our findings indicate that the $R$ polytypes (3$R$ and $R_{h}^{M}$), present when the twist angle between layers $\theta \rightarrow 0^\circ$, demonstrate higher diffusivity than the slightly more stable $H$ forms (present when $\theta \rightarrow 60^\circ$).

These discoveries hold significant importance for future studies concerning transport properties in materials with twisted layers, leading to moiré structures and the formation of differently sized domains with various high-symmetry stackings.
Such arrangements have the potential to form localized transport channels in these systems.

Furthermore, our results demonstrate the potential for optimizing intriguing materials based on TMDC homo- or hetero-layers with different twist angles for applications involving ion- or atom-transport, such as in fuel cells.

While our study focused on the diffusion of protiums, it is reasonable to expect that the results will be quite similar for protons as well.\cite{an2019chemistry}.

\section{Experimental Section}
\label{sec:methodology}
\textbf{Computational Details}

In this work, we investigated different semiconducting TMDCs from Group 6 (see Figure~\ref{fig1}), with a general formula MX$_2$ (M - Mo or W; X - S or Se).
MoS$_2$ was also simulated in the following high-symmetry stackings with two layers in the unit cell forming a bulk material: $H_h^h$ (also known as 2$H$ stacking), $R_h^M$ (also known as 3$R$ stacking if three layers are in the unit cell), $R_h^h$, $H_h^X$, and $H_h^M$.
For comparison, we have investigated two other layered materials, namely, metallic NbS$_2$ and semiconducting beta-InSe (see Figure~\ref{fig1}).
All systems were represented as bulk  materials and periodic boundary conditions were used accordingly.
To investigate the diffusion of H atom between layers of these materials, we used the 6$\times$6$\times$1 supercels for all studied systems.

All systems were fully relaxed (atomic positions and lattice vectors) using density functional theory (DFT)  with Perdew–Burke–Ernzerhof exchange-correlation functional (PBE),\cite{perdew1996generalized}
Gaussian augmented plane wave (GAPW) basis sets, and Grimme's D3 London dispersion correction,\cite{grimme2010consistent} as implemented in the CP2K package.\cite{hutter2014wiley}
The Quickstep method was employed with Goedecker–Teter–Hutter (GTH)\cite{goedecker1996separable} pseudopotentials together with DZVP-MOLOPT-GTH-SR basis set for all elements except hydrogen, which was represented using the full potential. We considered NbS$_2$ as a ferromagnetic metallic system.
In all calculations performed using supercells, we used $\Gamma$-point approach.
Since we were interested only in the free-energy barriers and the corresponding self-diffusion coefficients, we omitted using spin-orbit coupling, which is important when investigating electronic properties.

We performed well-tempered metadynamics (WTMetaD)\cite{barducci2008well} simulations, as implemented in the CP2K package, to obtain the free-energy barriers ($\Delta F$) for H-atom diffusion between the layers of studied materials. Here, the canonical NVT ensemble was employed with the temperature set to 300~K and CSVR (canonical sampling velocity rescaling) thermostat\cite{bussi2007canonical} with 0.5~fs timestep.
Each trajectory was at least 45 ps long, to ensure convergence.
Each WTMetaD simulation was preceded by standard Born-Oppenheimer molecular dynamics (BOMD) simulations (MD; NVT, 300 K, time step of 0.5 fs) with duration of at least 5 ps to ensure thermal equilibrium.
We defined two collective variables (CVs; see Figure~\ref{fig2}) to trigger the process of H-atom transfer between the layers:
all chalcogen atoms in a given system were divided into three different kinds: X1 (inner layer with bound H atom), X2 (inner layer neighboring X1), and X (outermost layers). One of the collective variables, CV1, was defined as the coordination number (CN) of H to X1, $CN_{H-X1}$, and the other, CV2, as CN of H to X2, $CN_{H-X2}$. The form of the coordination function, $CN_{H-X}$, is defined in CP2K as follows:

\begin{equation}\label{eq1}
\begin{split}
     CN_{H-X} = \sum^{H}_{i} \sum^{X}_{j} \frac{1 - (\frac{r_{ij}}{R_0})^{6}} {1 - (\frac{r_{ij}}{R_0})^{12}}\\
\end{split}
\end{equation}

\noindent where r$_{ij}$ is the distance between two sites that bind the H atom.
R$_0$, the cutoff distance in the $CN$, is set to 3.0~a$_0$ for X1-H and X2-H distances in all considered materials.\cite{an2019chemistry}
Note that we only considered transfer of H atoms between X1 and X2 because $\Delta F$ for transfer within the same layer is much larger.\cite{an2019chemistry}

\begin{figure}[h!]
 \includegraphics[width=0.75\columnwidth]{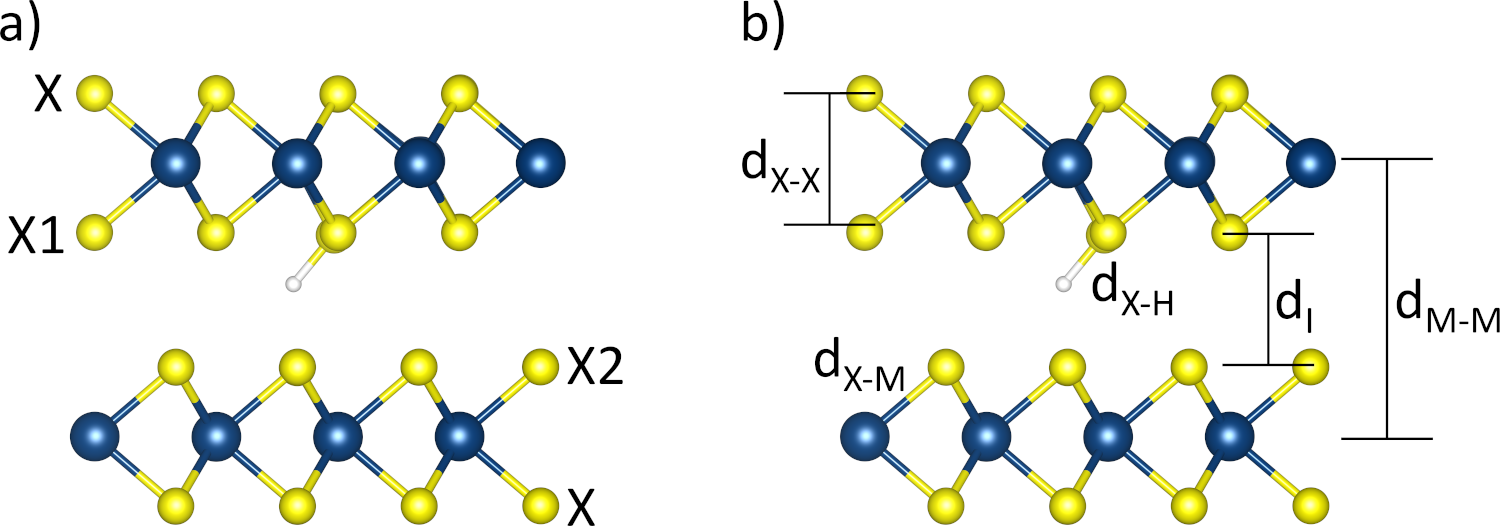}%
 \caption{\label{fig2} 
(a) An exemplary representation of collective variables X1-H and X2-H (see details in Methods section) within a TMDC material, shown from the side view. (b) Definition of selected bond lengths and interatomic distances (cf. Table~\ref{tab:table1}).}
\end{figure}

\noindent Gaussian hills were applied every 150 steps to all TMDCs and every 100 steps to InSe during WTMetaD calculations.
From WTMetaD simulations, we obtained the free-energy barriers for H atom transfer between layers, $\Delta F$, which was used to calculate the self-diffusion coefficients, $D$:\cite{beyer1982determination,an2019chemistry}

\begin{equation}\label{eq2}
\begin{split}
     D = D_{0}\exp{(\frac{\Delta F}{k_{B}T})} \\
     D_{0} = \frac{ (\Delta r)^{2} v_{0}}{q_{i}} 
\end{split}
\end{equation}

\noindent where $k_{B}$ is the Boltzmann's constant, T is the temperature (here set to 300~K), $D_{0}$ is a pre-factor,\cite{an2019chemistry,herrero2010diffusion} $\Delta r$ is the distance between H atoms in two closest binding sites, $v_{0}$ is the frequency of the stretching modes between X and H atoms, and $q_{i}$ can be 2, 4 or 6 for the diffusion dimensionality of 1D, 2D or 3D, respectively (here set to $q_{i}$ = 4). The frequency, $v_{0}$, of S-H and Se-H were taken from literature to be 2570 and 2300~cm$^{-1}$, respectively, assuming that the second neighbours do not affect it.\cite{freqs} 

\medskip
\textbf{Acknowledgements} \par

This research was supported by the Deutsche Forschungsgemeinschaft (projects GRK 2721/1 and SFB 1415). The authors acknowledge the high-performance computing center of ZIH Dresden, the Leipzig University Computing Centre, and the Paderborn Center for Parallel Computing (PC$^2$) for computational resources.
The authors thank Prof. Thomas Heine for fruitful discussions.
Language improvement assistance provided by ChatGPT (an AI language model developed by OpenAI) was utilized for enhancing the clarity and correctness of the manuscript.

\medskip

\newpage
\section{Supporting Information}
\begin{figure}[h!]
 \includegraphics[width=0.75\textwidth]{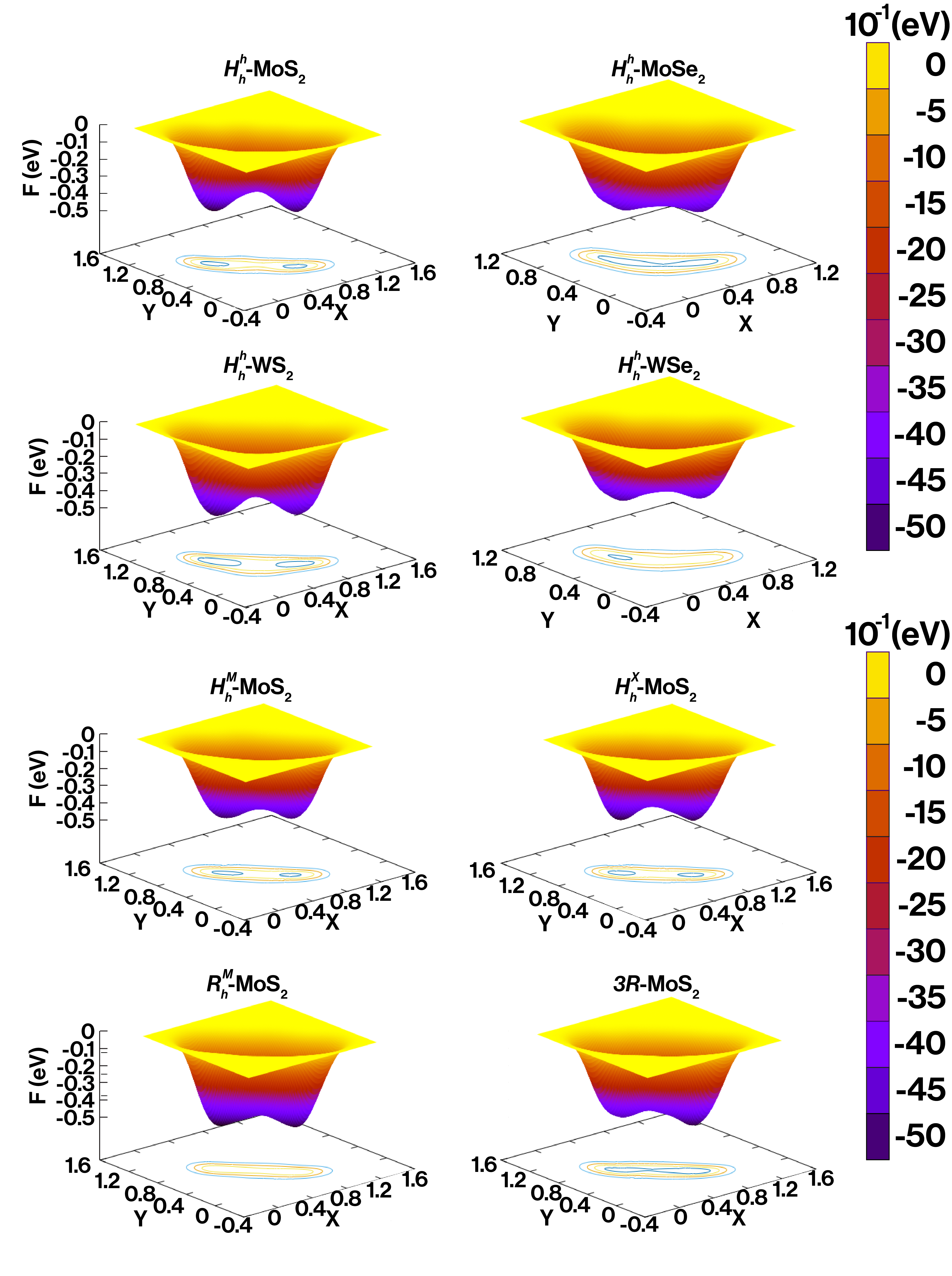}%
 \caption{\label{fig1} Free-energy ($F$) landscapes of all investigated systems obtained from WTMetaD simulations of a single H atom moving between layers of TMDC materials. The collective variables (CVs) were defined as coordination number (CN) of H atoms with the inner chalcogen atoms X1-H (CV1) and X2-H (CV2; cf. Figure 2 in main text), which correspond to the x and y axes. The color legend corresponds to $F$ in eV.}
\end{figure}

\begin{figure}[h!]
 \includegraphics[width=0.5\textwidth]{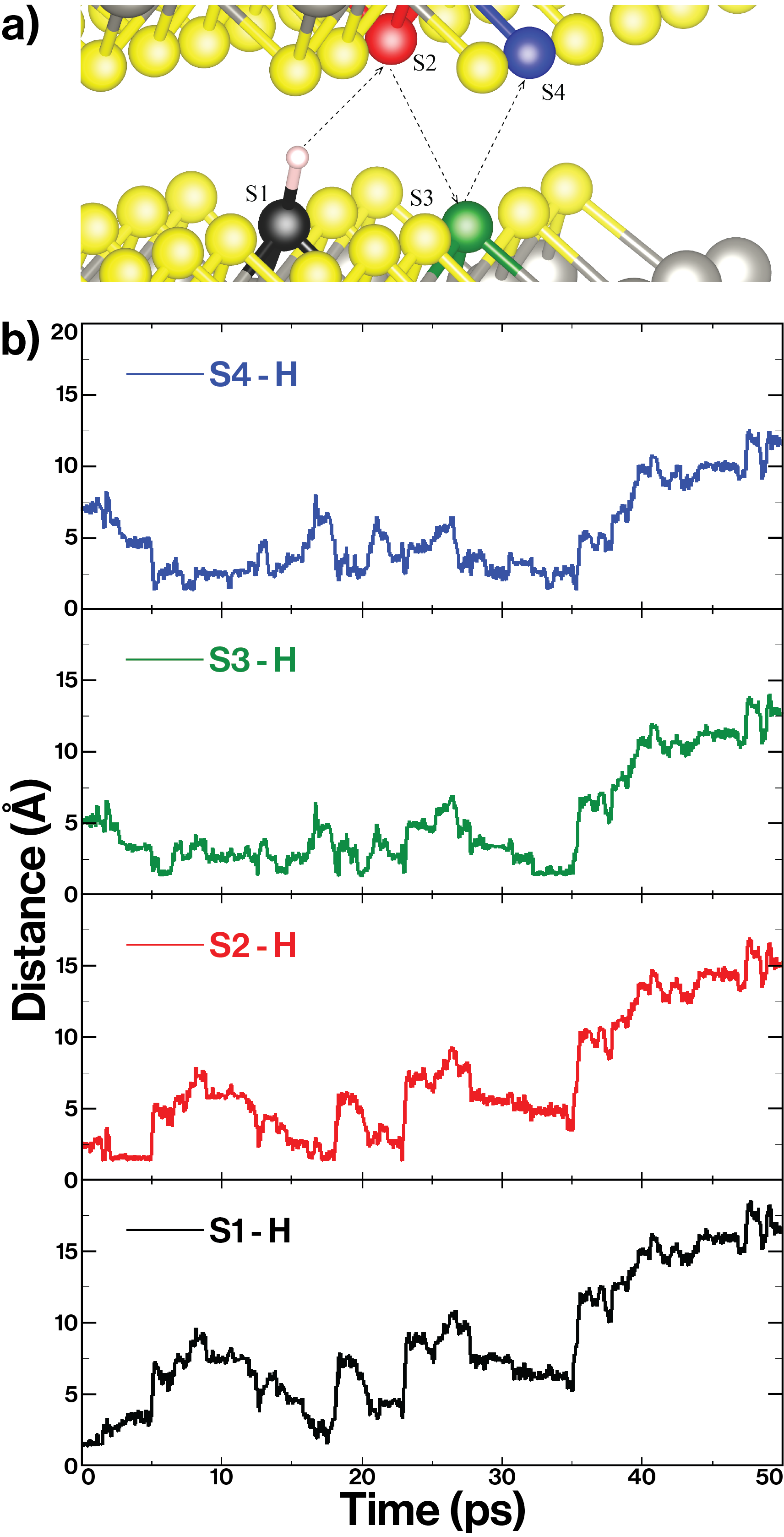}%
 \caption{\label{fig2} 
 (a) Schematics of the H-atom transfers path between layers of a TMDC material. (b) Exemplary X-H bond distance change during WTMetaD simulations in WS$_2$ bulk.}
\end{figure}

\end{document}